\documentclass[a4paper,12pt]{article}
\usepackage{graphicx}
\usepackage{epsfig}
\usepackage[american]{babel}

\begin{document}

\title{The effect of noise on the dynamics of a complex map at the period-tripling accumulation point}
\author{Olga~B.~Isaeva, Sergey~P.~Kuznetsov \\
and \\ Andrew~H.~Osbaldestin}
\date{}
\maketitle\begin{center} \emph{Institute of Radio-Engineering and
Electronics of RAS, Saratov Branch, \\ Zelenaya 38, Saratov,
410019, Russia}\end{center}

\maketitle\begin{center} \emph{Department of Mathematics,
University of Portsmouth,
\\ Portsmouth, PO1 3HE, UK}\end{center}

\begin{abstract}
As shown recently (O.B.Isaeva et al., Phys.Rev E64, 055201), the
phenomena intrinsic to dynamics of complex analytic maps under
appropriate conditions may occur in physical systems.  We study
scaling regularities associated with the effect of additive noise
upon the period-tripling bifurcation cascade generalizing the
renormalization group approach of Crutchfield et al.
(Phys.Rev.Lett., 46, 933) and Shraiman et al. (Phys.Rev.Lett., 46,
935), originally developed for the period doubling transition to
chaos in the presence of noise. The universal constant determining
the rescaling rule for the intensity of the noise in
period-tripling is found to be $\gamma=12.2066409...$ Numerical
evidence of the expected scaling is demonstrated.

\end{abstract}

\maketitle

\section{INTRODUCTION}
Application of the renormalization group (RG) analysis and the
concepts of universality and scaling in nonlinear dynamics started
with the works of Feigenbaum concerning the period-doubling
transition to chaos~\cite{1}. The universal quantitative
regularities intrinsic to this type of behavior are common for a
wide class of systems including one-dimensional maps, forced
dissipative nonlinear oscillators, R\"{o}ssler and Lorenz models,
experimental systems in hydrodynamics, electronics, laser physics,
etc~\cite{2}. In the context of real systems, a question of
vital importance is the effect of noise on the dynamics at the onset of
chaos. It is more or less obvious that the presence of noise
destroys subtle details of the small-scale (or long-time)
dynamics. The RG analysis developed by Crutchfield et al.~\cite{3}
and by Shraiman et al. ~\cite{4} gives a quantitative measure for
this effect. Namely, to observe one more level of period
doubling it is needed to decrease the noise amplitude by the
factor $\gamma=6.61903$.

After Feigenbaum the RG approach was developed by many authors,
e.g. for the onset of chaos via quasiperiodicity and
intermittency~\cite{5,6}, as well as for some situations arising
in the multiparameter analysis of transition to chaos~\cite{7} or
to strange nonchaotic attractors~\cite{8}. In particular, the
effects of noise on the dynamics have been studied in dissipative
systems for intermittency~\cite{9}, quasiperiodicity~\cite{10},
bicritical behavior~\cite{11}, and in Hamiltonian systems for
period doubling~\cite{12} and KAM-torus destruction~\cite{13}.

A special kind of dynamics occurs in iterative complex analytic
maps $z_{n+1}=f(z_{n})$~\cite{14}. An example is a complex
quadratic map
\begin{equation} \label{1}
z_{n+1} = \lambda - z_{n}^{2},
\end{equation}
which gives rise to remarkable fractal formations in the complex
plane of the variable $z$ (the Julia sets) and of the complex
parameter $\lambda$ (the Mandelbrot set). The latter is defined by
the condition that being launched from the origin, the iterations
at a given $\lambda$ never diverge. The Mandelbrot set has a
subtle and complicated structure, which is a subject of extensive
research. In particular, it contains special points of
accumulation of bifurcation cascades. Beside the period doubling,
there are also cascades of period tripling, quadrupling, etc. The
period-tripling accumulation point has been studied first by
Golberg, Sinai, and Khanin~\cite{15} (see also~\cite{16}), and
will be referred to as the \it GSK critical point \rm. In the
map~(\ref{1}) it is located at
\begin{equation} \label{2}
\lambda_{c}=0.0236411685377 + 0.7836606508052 i.
\end{equation}
By analogy with Feigenbaum universality, it could be expected to
occur in other nonlinear systems as well.

However, in fact, complex analytic functions represent a very
special and restricted class of maps because real and imaginary
parts of $f(z)$ must satisfy the Cauchy-Riemann equations. If this
is not the case, the dynamics become drastically
different~\cite{17}. In particular, it was shown that the
period-tripling type of behavior does not survive, in general,
under a non-analytic perturbation~\cite{18} (in spite of the
claim in the original paper~\cite{15} and some other
works~\cite{19,20}). So, a principal question is: Do the phenomena
intrinsic to complex analytic maps have any concern to
dynamical behavior of physical systems? One example of an
appropriate physical system was suggested by Beck~\cite{21}.
Another approach developed in Ref.~\cite{22} is based on a
construction of a system of two coupled units, each of which can
demonstrate Feigenbaum's period-doubling cascade, and the coupling
is of some special kind. Moreover, this idea has been verified in
an experiment with coupled electronic circuits.

Recognizing a possibility of the physical occurrence of the phenomena
of complex analytic dynamics, we intend to study in this paper the
effect of noise on the period-tripling cascade in the spirit of
the earlier works of Crutchfield et al. ~\cite{3} and Shraiman et
al. ~\cite{4}, on a basis of the appropriate generalization of the
RG approach.

\section{RENORMALIZATION GROUP ANALYSIS}
If we perform $3^{k}$ iterations of the map~(\ref{1}) at the GSK
critical point $\lambda=\lambda_{c}$, the renormalized evolution
operator will converge, as known, to a universal function
\begin{equation}\label{3}
\lim \limits_{k \rightarrow
\infty}f^{3k}(zf^{3k}(0))/f^{3k}(0)=g(z),
\end{equation}
which satisfies the fixed-point RG equation
\begin{equation}\label{4}
g(z)=\alpha g(g(g(z/\alpha))).
\end{equation}
Obviously, $g(z)$ represents an evolution operator at the critical
point for an asymptotically large number of iterations $3^{k}$ in
terms of the properly normalized dynamical variable. Numerically,
the approximation for this function was found in Ref.~\cite{15} as
a finite Taylor series:
\begin{equation}\label{5}
g(z)\cong 1+(0.0547-0.7490 i)z^{2}+(-0.0244-0.0525 i)z^{4}+...
\end{equation}
together with the complex scaling constant
\begin{equation}\label{6}
\alpha=1/g(g(g(0)))\cong -2.09691989+2.35827964 i.
\end{equation}

Let us introduce noise and consider a stochastic equation
\begin{equation}\label{7}
z_{n+1}=g(z_{n})+\varepsilon\varphi(z_{n})\xi_{n},
\end{equation}
where $\varepsilon$ is a small parameter of the noise intensity,
$\varphi(z)$ is a smooth real function of complex argument, and
$\xi_{n}$ is a complex stationary random sequence with
statistically independent subsequent terms. We assume that it has
zero mean, unit mean square $\langle|\xi_{n}^{2}|\rangle=1$, and
zero correlation of the real and imaginary parts,
$\langle(\mathrm{Re}\xi_{n})(\mathrm{Im}\xi_{n})\rangle=0$. It is
expected that the scaling properties of the response under study
will be independent on the concrete form of the distribution
function, although it is convenient to suppose that a distribution
function for $\xi_{n}$ is of such kind that the amplitude of the
noise is bounded.

Three-fold application of the stochastic map yields, in the first
order in $\varepsilon$,
\begin{equation}\label{8}\begin{array}{l}
z_{n+3}=g(g(g(z_{n})))
+\varepsilon\{g'(g(g(z_{n})))g'(g(z_{n}))\varphi(z_{n})\xi_{n} \\+
g'(g(g(z_{n})))\varphi(g(z_{n}))\xi_{n+1}+\varphi(g(g(z_{n})))\xi_{n+2}\}.
\end{array}\end{equation}
Now, we renormalize the variable $z$ by substitution $z\rightarrow
z/\alpha$, and obtain
\begin{equation}\label{9}\begin{array}{l}
z_{n+3}=\alpha g(g(g(z_{n}/\alpha)))
  +\varepsilon\alpha\{g'(g(g(z_{n}/\alpha)))g'(g(z_{n}/\alpha))\varphi(z_{n}/\alpha)\xi_{n} \\ +
g'(g(g(z_{n}/\alpha)))\varphi(g(z_{n}/\alpha))\xi_{n+1}+\varphi(g(g(z_{n}/\alpha)))\xi_{n+2}\}.
\end{array}\end{equation}
The first term in the right-hand part of the equation equals
$g(z)$, in accordance with Eq.~(\ref{4}). Concerning the remaining
terms, we make the following important remark. Let us suppose that
we start at some $z_{n}.$ Consider an ensemble of the random
numbers $\{\xi_{n}, \xi_{n+1}, \xi_{n+2}\}$ and compose them with
complex coefficients given by functions of $z_{n}.$ As $\{\xi_{n},
\xi_{n+1}, \xi_{n+2}\}$ are independent, the sum can be
represented again as a random complex number with zero mean and
unit mean square multiplied by a real function of complex
argument, namely,
\begin{equation}\label{10}\begin{array}{l}
  \alpha\{g'(g(g(z_{n}/\alpha)))g'(g(z_{n}/\alpha))\varphi(z_{n}/\alpha)\xi_{n}\\ \qquad\quad\quad+
g'(g(g(z_{n}/\alpha)))\varphi(g(z_{n}/\alpha))\xi_{n+1}
+\varphi(g(g(z_{n}/\alpha)))\xi_{n+2}\}=\hat{\varphi}(z_{n})\hat{\xi}_{n}.
\end{array}\end{equation}
So, we rewrite~(\ref{9}) in the form analogous to~(\ref{7}), with
redefined function and random variable in the stochastic term:
\begin{equation}\label{11}
z_{n+3}=g(z_{n})+\varepsilon \hat{\varphi}(z_{n})\hat{\xi}_{n}.
\end{equation}
To obtain a closed functional equation, we multiply both parts of
Eq.~(\ref{10}) by the complex conjugates, and perform an averaging
over an ensemble of realizations of the noise. As
$\langle|\hat{\xi}^{2}_{n}|\rangle=\langle|\xi^{2}_{n}|\rangle=1$,
and
$\langle\xi_{n}\xi^{*}_{n+1}\rangle=\langle\xi_{n}\xi^{*}_{n+2}\rangle=0$,
we come to the relation
\begin{equation}\label{12}\begin{array}{l}
\hat{\Phi}(z_{n})=
|\alpha|^{2}\{|g'(g(g(z_{n}/\alpha)))g'(g(z_{n}/\alpha))|^{2}\Phi(z_{n}/\alpha)+
|g'(g(g(z_{n}/\alpha)))|^{2}\Phi(g(z_{n}/\alpha))+\Phi(g(g(z_{n}/\alpha)))\},
\end{array}\end{equation}
where $\Phi(z)=[\varphi(z)]^{2}$. Obviously, Eq.~(\ref{12}) has a
structure $\hat{\Phi}(z)=\mathbf{L}\Phi(z)$, where $\mathbf{L}$ is
a linear operator of the functional transformation given by the
right-hand part of~(\ref{12}). Repetitive application of the same
procedure to Eq.~(\ref{11}) yields a sequence of functions with
asymptotic behavior $\Phi_{k}(z)\cong\Omega^{k}\Phi(z)$,
determined by eigenvector $\Phi(z)$, associated with the largest
eigenvalue $\Omega$ of the operator $\mathbf{L}$:
\begin{equation}\label{13}\begin{array}{l}
\Omega\Phi(z)=
|\alpha|^{2}\{|g'(g(g(z/\alpha)))g'(g(z/\alpha))|^{2}\Phi(z/\alpha)+
|g'(g(g(z/\alpha)))|^{2}\Phi(g(z/\alpha))+\Phi(g(g(z/\alpha)))\}.
\end{array}\end{equation}

As mentioned above, the universal function $g(z)$ has been
approximated as a finite expansion over even powers of the
argument~\cite{15}. Using this data, we have numerically
constructed the functional transformation of the right-hand part
of Eq~(\ref{13}). The unknown function $\Phi(z)$ is represented by
a set of real values at nodes of a grid in a square $\{-1.2
<\mathrm{Re} z, \mathrm{Im} z <1.2\}$, and by an interpolation
scheme between them. Taking random initial conditions for
$\Phi(z)$, we perform the functional transformation and normalize
the resulting function as $\Phi^{0}(z)= \Phi(z)/\Phi(0)$. This
operation is repeated many times, until the form of the function
stabilizes (see Fig.~1). The value of $\Phi(0)$. (before the
normalization) converges to the eigenvalue $\Omega=149.0020828$.

\begin{figure}
\centerline{\includegraphics[width=0.4\textwidth,keepaspectratio]{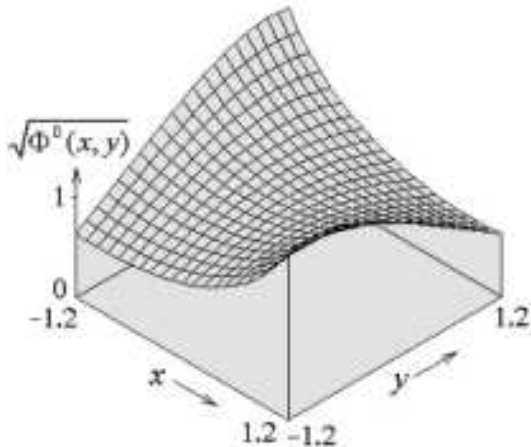}}
\caption{Plot of the eigenfunction responsible for the effect of
noise at the period tripling accumulation point obtained
numerically (see text).}
\end{figure}

In the linear approximation with respect to the noise amplitude,
the stochastic map for the evolution over $3^{k}$ steps at the GSK
critical point asymptotically may be written as
\begin{equation}\label{14}
z_{n+3^{k}}=g(z_{n})+\varepsilon\gamma^{k}\varphi(z_{n})\xi_{n},
\end{equation}
where
\begin{equation}\label{15}
\varphi(z)=\sqrt{\Phi^{0}(z)},\quad
\gamma=\sqrt{\Omega}=12.20664093.
\end{equation}

If we consider a small shift of $\lambda$ from the GSK critical
point, then an additional perturbation term appears in the
equation:
\begin{equation}\label{16}
z_{n+3^{k}}=g(z_{n})+C_{1}\delta_{1}^{k}h_{1}(z_{n})+\varepsilon\gamma^{k}\varphi(z_{n})\xi_{n}.
\end{equation}
Here, $h_{1}(z)$ represents an eigenvector of the linearized RG
equation without noise associated with the eigenvalue
$\delta_{1}=4.60022558-8.98122473i$. The coefficient $C_{1}$
depends on the parameter and vanishes at the critical point GSK.
In a close neighborhood of the critical point it is sufficient to
consider only the leading terms of the expansion and set
$C_{1}\propto (\lambda-\lambda_{c})$.

Now, we are ready to formulate the basic scaling property that
follows from~(\ref{16}).

If we triple the number of time steps (i.e., change $k$ to $k+1$),
decrease the parameter difference
$\Delta\lambda=\lambda-\lambda_{c}$ by division by $\delta_{1}$,
and decrease noise amplitude $\varepsilon$ by factor $\gamma$,
then the form of the stochastic map~(\ref{16}) remains unchanged.
Thus, with the new parameters, ($\Delta\lambda/ \delta_{1}$,
$\varepsilon/\gamma$) the noisy system will demonstrate the same
behavior as that with the old ones, but with a tripled time scale.

\section{MODEL MAP AND NUMERICAL EXPERIMENTS}
To verify the RG results in numerical experiments, we use a model
map with additive noise:
\begin{equation}\label{17}
z_{n+1}=\lambda-z_{n}^{2}+\varepsilon\xi_{n}.
\end{equation}
Real and imaginary parts for each term of the complex random
sequence $\xi_{n}$, are obtained as sums of $N=10$ zero-mean
computer generated pseudo-random numbers, which are properly
normalized to have $\langle|\xi_{n}^{2}|\rangle=1$. (In accordance
with the central limit theorem, $\xi_{n}$ is very close to a
Gaussian random variable.)

Fig.~2 illustrates fine structure of the Mandelbrot set near the
period-tripling accumulation point as it looks~(a) in the absence
and~(b) in the presence of noise. The pictures represent the
Lyapunov charts for the complex parameter $\lambda$.
(See~\cite{23,18} for previous applications of this method.) At
each pixel of the two-dimensional plot we estimate the Lyapunov
exponent
\begin{equation}\label{18}
\Lambda=\lim\limits_{N\rightarrow\infty}N^{-1}\sum\limits_{n=1}^{N}\ln|2z_{n}|
\end{equation}
from numerical computations, and mark the pixel with a gray tone.
The tones vary from dark to light as the Lyapunov exponent varies
from large negative to zero: white corresponds to zero, and black
to positive Lyapunov exponent. Divergence is shown by uniform
coloring with one special gray tone. The GSK critical point is
located exactly at the center of the diagram. A small box
containing the critical point, is magnified and rotated in
accordance with multiplication by $\delta_{1} \approx
4.6002-8.9812i$. As the parameter rescaling is associated with a
tripling of characteristic time scale, we accompany it with
reducing an interval for gray coding of Lyapunov exponents by
factor $1/3$ for each successive picture in the row. On the
panel~(a) such diagrams are shown for zero amplitude of
noise, $\varepsilon=0$, and on panel~(b) for noise of
fixed intensity, $\varepsilon=0.001$. Observe the similarity of
the leaves of the "Mandelbrot cactus" in the first case, and the
progressive destruction of the structure at subsequent levels
of the resolution in the second.

In Fig.~3
\begin{figure}
\centerline{\includegraphics[width=0.75\textwidth,keepaspectratio]{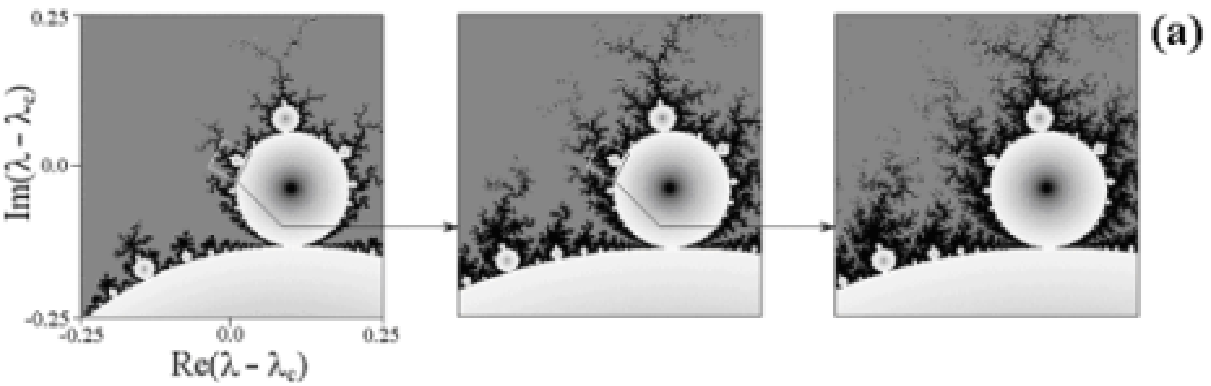}}
\centerline{\includegraphics[width=0.75\textwidth,keepaspectratio]{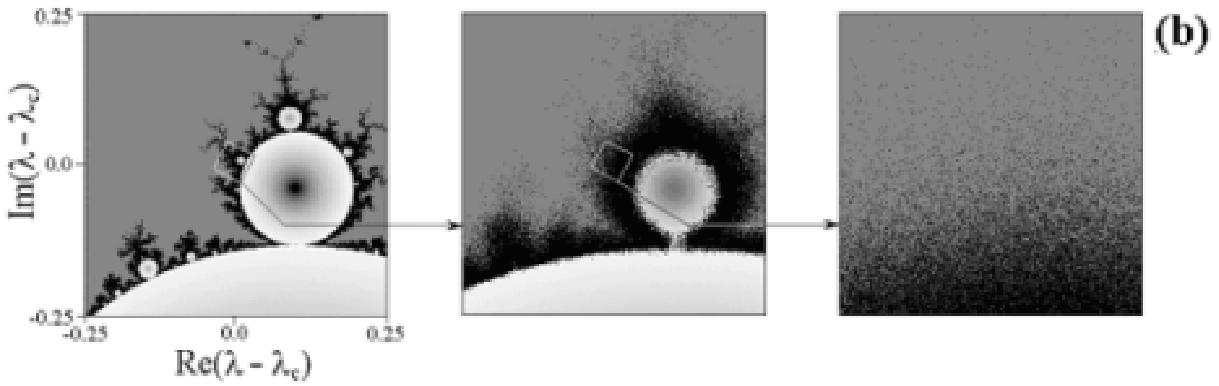}}

\caption{The Lyapunov charts for the model map~(\ref{17}) on the
complex parameter plane for zero intensity of noise
$\varepsilon=0$~(a) and in presence of noise,
$\varepsilon=0.001$~(b). The GSK critical point is located exactly
at the center of the diagrams. The tones vary from dark to light
as the Lyapunov exponent varies from large negative to zero: white
corresponds to zero, and black to positive values. Divergence is
shown by uniform coloring with one special gray tone. A small box
containing the critical point, is magnified and rotated in
accordance with multiplication by $\delta_{1}\approx
4.6002-8.9812i$. The interval for coded values of Lyapunov
exponents is reduced with factor $1/3$ for each successive picture
in the row to visualize self-similarity of the pictures in absence
of noise.}

\hspace{20mm}

\centerline{\includegraphics[width=0.75\textwidth,keepaspectratio]{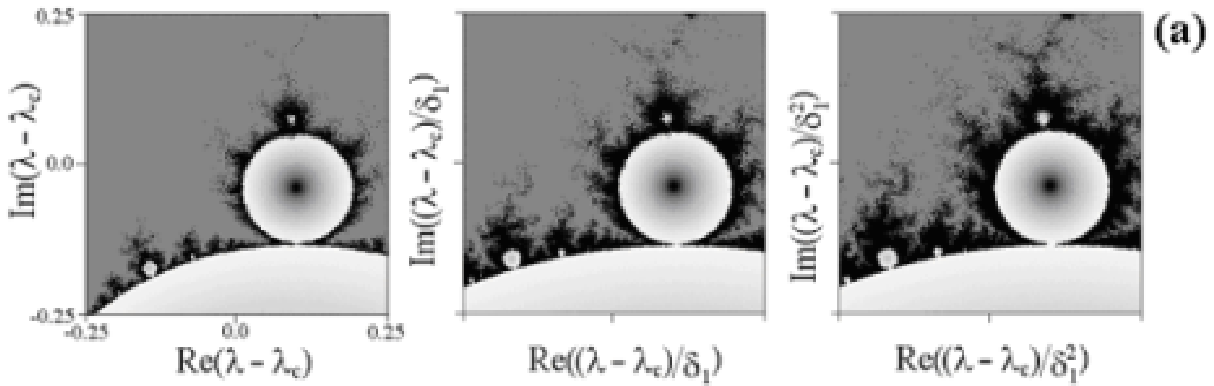}}
\centerline{\includegraphics[width=0.75\textwidth,keepaspectratio]{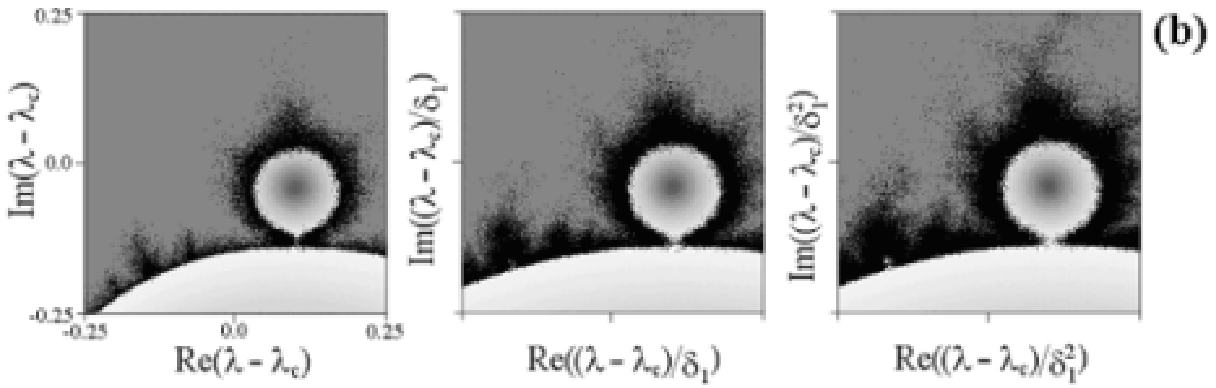}}

\caption{Lyapunov charts for the stochastic model~(\ref{17}) on
the complex parameter plane in presence of noise for
$\varepsilon=\varepsilon_{1}=0.003$,
$\varepsilon=\varepsilon_{1}/\gamma$, and
$\varepsilon=\varepsilon_{1}/\gamma^{2}$~(a), and for
$\varepsilon=\varepsilon_{2}=0.009$,
$\varepsilon=\varepsilon_{2}/\gamma$, and
$\varepsilon=\varepsilon_{2}/\gamma^{2}$~(b). The GSK critical
point is located exactly at the center of the diagrams. The
coordinate rescaling rule and the gray coding are the same as in
Fig.~2. The interval for coded values of Lyapunov exponents is
reduced with factor $1/3$ for each succesive picture in the row to
visualize the similarity of the pictures.}
\end{figure}
we show analogous pictures with noise, but now its intensity is
reduced for each subsequent picture in a row by the factor
$\gamma=12.2066...$ found from the RG analysis.
Panels~(a) and~(b) correspond to different
initial levels of the noise, respectively, $\varepsilon=0.003$ and
$0.009$. Now, the similarity of the pictures is restored. With
noise reduction by $\gamma$, just one more level of smaller leaves
of the Mandelbrot cactus reveals itself near the GSK critical
point.

Fig.~4
\begin{figure}
\centerline{\includegraphics[width=0.39\textwidth,keepaspectratio]{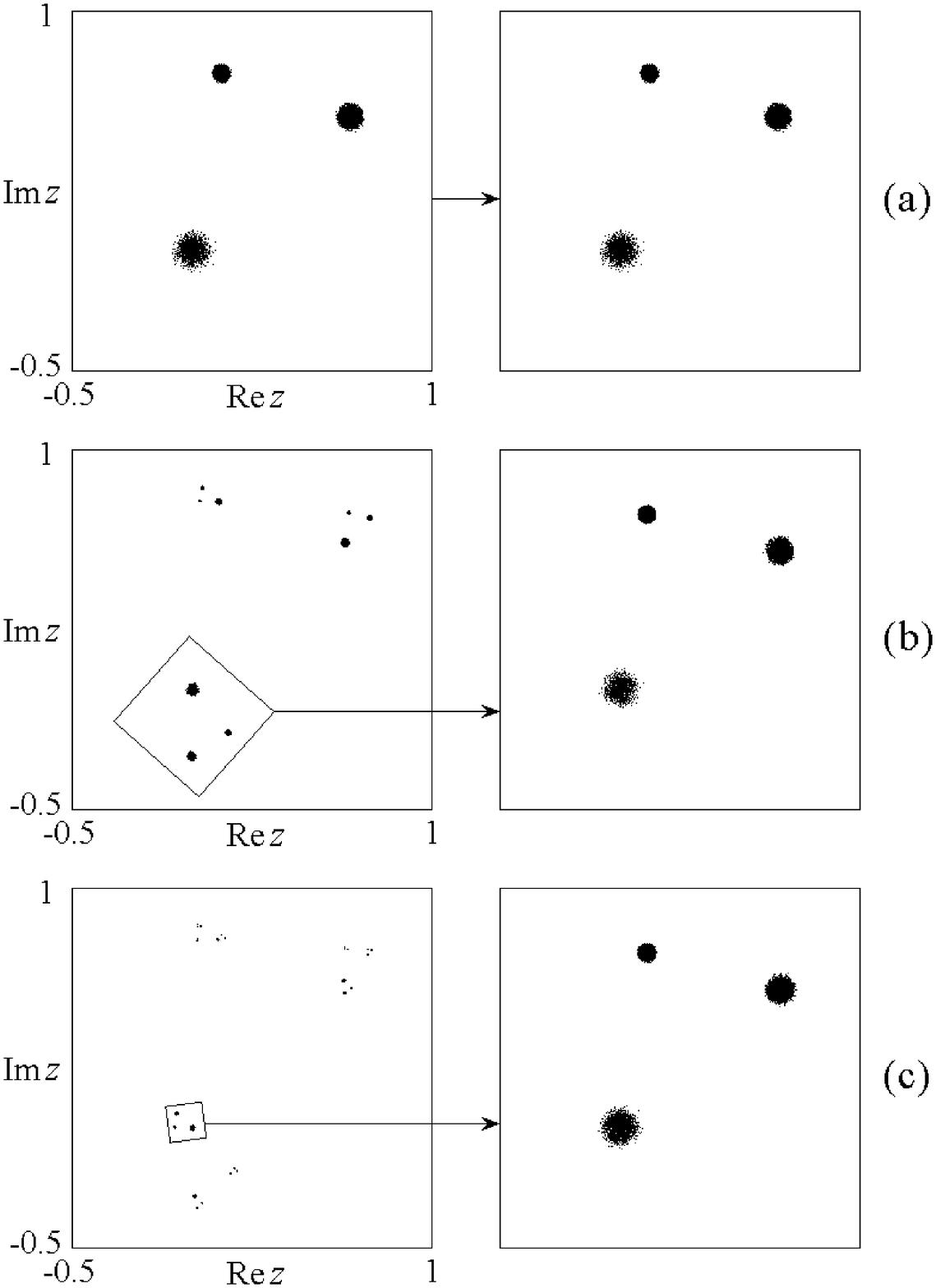}}
\caption{Similarity in structure of the noisy attractors for the
model map~(\ref{17}) on the plane of complex variable $z$. The
parameter $\lambda$ values are $0.122561167+0.744861767i$~(a),
$0.031552975+0.790783175i$~(b), $0.023369416+0.784679678i$~(c),
and the noise intensities are $\varepsilon=0.015$, $0.015/\gamma$,
and $0.015/\gamma^2$, respectively, where $\gamma=12.2066…$ The
right-hand pictures represent the definite parts of the attractors
drawn with rescaling by complex factors
$\alpha=-2.0959...+i2.3582...$~(b) and $\alpha^2$~(c).}
\end{figure}
illustrates similarity in structure of the noisy attractors for
the model map~(\ref{17}) in the plane of the complex variable $z$.
Diagrams~(b) and~(c) relate to noise intensities
reduced by the factors $\gamma$ and $\gamma^{2}$ in comparison
with the plot~(a). The parameter $\lambda$ values
correspond to the superstable cycles of period $3$, $9$, and $27$,
respectively. The right-hand panels show the indicated parts
of the pictures drawn with rescaling by complex factors $\alpha$
(diagram~(b)) and $\alpha^{2}$ (diagram~(c)).
Observe the similarity of the right-hand diagrams.

\section{CONCLUSION}
In this paper we have studied the effect of noise on a complex
analytic map at the period-tripling accumulation point. It was
shown that the effect of noise obeys regularities of a similar
nature to those reported for period doubling in real
one-dimensional maps. Namely, to observe one more level of the
bifurcation cascade it is needed to decrease the intensity of
noise by some definite scaling factor. For the case of the
period-tripling this factor has been found to be $12.2066...$, as
follows from the renormalization group analysis.
Also we have demonstrated
the scaling properties associated with the new constant in
numerical experiments.

The undertaken analysis is essential for discussion of the
possibility of observation of phenomena of complex analytic
dynamics in real physical experiment, e.g. in mechanics or
electronics~\cite{21,22,18}. The presence of noise in such systems
is inevitable, and our result gives a quantitative foundation for
estimates of an observable number of levels in the bifurcation
cascade.

Our work puts the phenomena in complex analytic map into the same
context in respect to the effect of noise, as other situations of
universal scaling behavior: period doubling~\cite{3,4},
intermittency~\cite{9}, quasiperiodicity~\cite{10},
bicriticality~\cite{11}, scaling phenomena in Hamiltonian
systems~\cite{12,13}. So, this specific field is enriched with one
more nontrivial example of the scaling behavior linked with
presence of noise.

The approach developed in this article may be adapted to study
the effect of noise on other
bifurcation cascades in complex analytic maps ("$n$-tupling" with
$n=4,5,...$~\cite{13,16}). Moreover, it gives a basis from which
to pose more
general questions on the effect of noise on the Mandelbrot-like
sets in physical systems possessing such kind of objects in their
parameter space.


\section*{ACKNOWLEDGEMENT}
The authors acknowledge support from UK Royal Society. OBI and SPK
thank Research Educational Center of Nonlinear Dynamics and
Biophysics at Saratov State University (REC-006), and RFBR for
support under grant No 03-02-16092.

\end{document}